\documentclass[12pt]{article}

\usepackage{axodraw}
\usepackage{epsfig}

\makeatletter

\def\paragraph{\@startsection{paragraph}{4}{\z@}{+2.00ex plus
 +1ex minus +.2ex}{1.5ex plus .2ex}{\it\normalsize}}

\def\section{\@startsection {section}{1}{\z@}{+3.0ex plus +1ex minus
  +.2ex}{2.3ex plus .2ex}{\normalsize\bf\boldmath}}
\def\subsection{\@startsection{subsection}{2}{\z@}{+2.5ex plus +1ex
minus +.2ex}{1.5ex plus .2ex}{\normalsize\bf\boldmath}}
\def\subsubsection{\@startsection{subsubsection}{3}{\z@}{+3.25ex plus
 +1ex minus +.2ex}{1.5ex plus .2ex}{\normalsize\it}}

\expandafter\ifx\csname mathrm\endcsname\relax\def\mathrm#1{{\rm #1}}\fi

\newcounter{saveeqn}

\@addtoreset{equation}{section}

\newcount\@tempcntc
\def\@citex[#1]#2{\if@filesw\immediate\write\@auxout{\string\citation{#2}}\fi
  \@tempcnta\z@\@tempcntb\m@ne\def\@citea{}\@cite{\@for\@citeb:=#2\do
    {\@ifundefined
       {b@\@citeb}{\@citeo\@tempcntb\m@ne\@citea
        \def\@citea{,\penalty\@m\ }{\bf ?}\@warning
       {Citation `\@citeb' on page \thepage \space undefined}}%
    {\setbox\z@\hbox{\global\@tempcntc0\csname
b@\@citeb\endcsname\relax}%
     \ifnum\@tempcntc=\z@ \@citeo\@tempcntb\m@ne
       \@citea\def\@citea{,\penalty\@m}
       \hbox{\csname b@\@citeb\endcsname}%
     \else
      \advance\@tempcntb\@ne
      \ifnum\@tempcntb=\@tempcntc
      \else\advance\@tempcntb\m@ne\@citeo
      \@tempcnta\@tempcntc\@tempcntb\@tempcntc\fi\fi}}\@citeo}{#1}}

\def\@citeo{\ifnum\@tempcnta>\@tempcntb\else\@citea
  \def\@citea{,\penalty\@m}%
  \ifnum\@tempcnta=\@tempcntb\the\@tempcnta\else
   {\advance\@tempcnta\@ne\ifnum\@tempcnta=\@tempcntb \else
\def\@citea{--}\fi
    \advance\@tempcnta\m@ne\the\@tempcnta\@citea\the\@tempcntb}\fi\fi}

\def\nl{\nonumber\\}

\def\asymp#1%
{\mathrel{\raisebox{-.4em}{$\widetilde{\scriptstyle #1}$}}}

\def\Nequal#1%
{\mathrel{\raisebox{-.5em}{$\stackrel{=}{\scriptstyle\rm#1}$}}}
\newcommand{\dsl}[1]{\not \hspace{-0.7mm}#1}
\def\dsl{\mathpalette\make@slash}
\def\make@slash#1#2{\setbox\z@\hbox{$#1#2$}%
  \hbox to 0pt{\hss$#1/$\hss\kern-\wd0}\box0}

\def\beq{\begin{equation}}
\def\eeq{\end{equation}}
\def\beqar{\begin{eqnarray}}
\def\eeqar{\end{eqnarray}}
\def\barr#1{\begin{array}{#1}}
\def\earr{\end{array}}
\def\bfi{\begin{figure}}
\def\efi{\end{figure}}
\def\btab{\begin{table}}
\def\etab{\end{table}}
\def\bce{\begin{center}}
\def\ece{\end{center}}

\def\text{\textstyle}

\def\al{\alpha}

\def\ga{\gamma}
\def\de{\delta}

\def\si{\sigma}

\def\reffi#1{\mbox{Figure~\ref{#1}}}
\def\reffis#1{\mbox{Figures~\ref{#1}}}
\def\refta#1{\mbox{Table~\ref{#1}}}
\def\reftas#1{\mbox{Tables~\ref{#1}}}

\def\citere#1{\mbox{Ref.~\cite{#1}}}
\def\citeres#1{\mbox{Refs.~\cite{#1}}}

\newcommand{\GeV}{\unskip\,\mathrm{GeV}}
\newcommand{\MeV}{\unskip\,\mathrm{MeV}}
\newcommand{\keV}{\unskip\,\mathrm{keV}}

\newcommand{\fb}{\unskip\,\mathrm{fb}}

\newcommand{\rd}{{\mathrm{d}}}

\newcommand{\Oa}{\mathswitch{{\cal{O}}(\alpha)}}

\def\mathswitchr#1{\relax\ifmmode{\mathrm{#1}}\else$\mathrm{#1}$\fi}

\newcommand{\PW}{\mathswitchr W}
\newcommand{\Pw}{\mathswitchr w}
\newcommand{\PZ}{\mathswitchr Z}

\newcommand{\PH}{\mathswitchr H}
\newcommand{\Pe}{\mathswitchr e}

\newcommand{\Pd}{\mathswitchr d}
\newcommand{\Pdbar}{\bar{\mathswitchr d}}
\newcommand{\Pu}{\mathswitchr u}
\newcommand{\Pubar}{\bar{\mathswitchr u}}
\newcommand{\Ps}{\mathswitchr s}

\newcommand{\Pcbar}{\bar{\mathswitchr c}}

\newcommand{\Pt}{\mathswitchr t}

\newcommand{\Pep}{\mathswitchr {e^+}}
\newcommand{\Pem}{\mathswitchr {e^-}}
\newcommand{\PWp}{\mathswitchr {W^+}}
\newcommand{\PWm}{\mathswitchr {W^-}}

\def\mathswitch#1{\relax\ifmmode#1\else$#1$\fi}

\newcommand{\MW}{\mathswitch {M_\PW}}
\newcommand{\MWfit}{\mathswitch {M_{\PW,\mathrm{fit}}}}

\newcommand{\MZ}{\mathswitch {M_\PZ}}
\newcommand{\MH}{\mathswitch {M_\PH}}
\newcommand{\Me}{\mathswitch {m_\Pe}}

\newcommand{\Mt}{\mathswitch {m_\Pt}}
\newcommand{\GW}{\Gamma_{\PW}}
\newcommand{\GWfit}{\mathswitch {\Gamma_{\PW,\mathrm{fit}}}}
\newcommand{\GZ}{\Gamma_{\PZ}}

\newcommand{\sw}{\mathswitch {s_\Pw}}
\newcommand{\cw}{\mathswitch {c_\Pw}}

\newcommand{\GF}{\mathswitch {G_\mu}}

\def\solid{\raise.9mm\hbox{\protect\rule{1.1cm}{.2mm}}}
\def\dash{\raise.9mm\hbox{\protect\rule{2mm}{.2mm}}\hspace*{1mm}}

\def\ie{i.e.\ }

\newcommand{\born}{{\mathrm{Born}}}
\newcommand{\corr}{{\mathrm{corr}}}

\newcommand{\eeWWffff}{\Pep\Pem\to\PW\PW\to 4f}
\newcommand{\eeWWffffg}{\eeWWffff\gamma}
\newcommand{\eeffff}{\Pep\Pem\to 4f}
\newcommand{\eeffffg}{\eeffff\ga}
\renewcommand{\O}{{\cal O}}

\def\draftdate{\relax}
\def\mda{\relax}
\def\mua{\relax}
\def\mla{\relax}
\def\draft{
\def\thtystars{******************************}
\def\sixtystars{\thtystars\thtystars}
\typeout{}
\typeout{\sixtystars**}
\typeout{* Draft mode!
         For final version remove \protect\draft\space in source file *}
\typeout{\sixtystars**}
\typeout{}
\def\draftdate{\today}
\def\mua{\marginpar[\boldmath\hfil$\uparrow$]%
                   {\boldmath$\uparrow$\hfil}%
                    \typeout{marginpar: $\uparrow$}\ignorespaces}
\def\mda{\marginpar[\boldmath\hfil$\downarrow$]%
                   {\boldmath$\downarrow$\hfil}%
                    \typeout{marginpar: $\downarrow$}\ignorespaces}
\def\mla{\marginpar[\boldmath\hfil$\rightarrow$]%
                   {\boldmath$\leftarrow $\hfil}%
                    \typeout{marginpar: $\leftrightarrow$}\ignorespaces}
\def\Mua{\marginpar[\boldmath\hfil$\Uparrow$]%
                   {\boldmath$\Uparrow$\hfil}%
                    \typeout{marginpar: $\uparrow$}\ignorespaces}
\def\Mda{\marginpar[\boldmath\hfil$\Downarrow$]%
                   {\boldmath$\Downarrow$\hfil}%
                    \typeout{marginpar: $\downarrow$}\ignorespaces}
\def\Mla{\marginpar[\boldmath\hfil$\Rightarrow$]%
                   {\boldmath$\Leftarrow $\hfil}%
                    \typeout{marginpar: $\leftrightarrow$}\ignorespaces}
\overfullrule 5pt
\oddsidemargin -15mm
\marginparwidth 29mm
}

\def\stars{\strut\leaders\hbox{*}\hfill\strut}
\def\starline{\hfil\strut\hfil\hbox to \textwidth {\stars}\hfil}

\begin{document}
\thispagestyle{empty}
\def\thefootnote{\fnsymbol{footnote}}
\setcounter{footnote}{1}
\null
\draftdate\hfill BI-TP 99/45 \\
\strut\hfill  LU-ITP 1999/020\\
\strut\hfill PSI-PR-99-29\\
\strut\hfill UR-1591\\
\strut\hfill hep-ph/9912261
\vfill
\begin{center}
{\Large \bf\boldmath
$\O(\alpha)$ corrections to $\Pep\Pem\to\PW\PW\to4\mbox{ fermions}
(+\ga)$:
\\[.5em]
first numerical results from {\sc RacoonWW}
\par} \vskip 2.5em
\vspace{1cm}

{\large
{\sc A.\ Denner$^1$, S.\ Dittmaier$^2$, M. Roth$^{3}$ and 
D.\ Wackeroth$^4$} } \\[1cm]

$^1$ {\it Paul-Scherrer-Institut\\
CH--5232 Villigen PSI, Switzerland} \\[0.5cm]

$^2$ {\it Theoretische Physik, Universit\"at Bielefeld \\
D-33615 Bielefeld, Germany}
\\[0.5cm]

$^3$ {\it Institut f\"ur Theoretische Physik, Universit\"at Leipzig\\
D-04109 Leipzig, Germany}
\\[0.5cm]

$^4$ {\it Department of Physics, University of Rochester\\
Rochester, NY 14627-0171, USA}
\par \vskip 1em
\end{center}\par
\vskip 2cm {\bf Abstract:} \par First numerical results of the Monte
Carlo generator {\sc RacoonWW} for $\eeWWffff(+\gamma)$ in the
electroweak Standard Model are presented.  This event generator is the
first one that includes $\O(\al)$ electroweak radiative corrections in
the double-pole approximation completely.  We briefly describe the
strategy of the calculation and give numerical results for total cross
sections, including CC03,
and various distributions.
\par
\vskip 1cm
\noindent
December 1999
\null
\setcounter{page}{0}
\clearpage
\def\thefootnote{\arabic{footnote}}
\setcounter{footnote}{0}

At present, the focus of Standard-Model (SM) tests lies on
\PW-boson-pair production at LEP2, \ie on the process $\eeWWffff$.
In order to match the experimental accuracy of roughly 1\%,
theoretical predictions for the cross sections of $\eeffff$ with a
precision at or below the per-cent level are needed. This requires to
include the complete set of lowest-order diagrams for $\eeffff$ and
the $\Oa$ corrections to the W-pair production channels $\eeWWffff$.
Since the impact of electroweak corrections grows with increasing
energy, this task is even more important for future linear colliders
with higher energy and luminosity.

In all regions of phase space where \PW-pair production dominates the
cross section for $\eeffff$, an expansion of the matrix element about
the poles of the resonant \PW~propagators provides a reasonable
approximation for the radiative corrections.  Neglecting corrections
to the non-doubly-resonant contributions leads to uncertainties of the
order $\al/\pi\times\GW/\MW\times\log(\ldots)\sim 0.1\%$ with respect
to the leading lowest-order contributions, where the logarithm
indicates possible logarithmic enhancements.  Thus, this so-called
double-pole approximation (DPA) should be sufficient for an accuracy
of $0.5\%$
for observables that are dominated by doubly-resonant diagrams.
Moreover, the DPA provides a gauge-invariant answer and allows us to
use the existing results for on-shell W-pair production
\cite{rcwprod1,rcwprod2} and \PW-boson decay
\cite{rcwdecay1,rcwdecay2}, as far as the virtual corrections are
concerned.

The DPA for $\Oa$ corrections to pair production of unstable particles
has already been used in the literature.  A possible strategy has been
proposed in \citere{Ae94}.  A Monte Carlo generator including the
corrections to the \PW-pair production subprocess and the
leading-logarithmic corrections to the \PW-boson decays has been
constructed \citeres{ja97,ja99}, 
but non-factorizable corrections and W-spin correlations have been
neglected there.
A first complete calculation of the
$\Oa$ corrections for off-shell \PW-pair production, including a
numerical study of leptonic final states, was presented in
\citere{Be98} using a semi-analytical approach, which is, however,
only applicable to ideal theoretical situations.

In this paper we present the first complete calculation of the $\Oa$
corrections for off-shell \PW-pair production in DPA that has been
implemented in a Monte Carlo generator,
which is called {\sc RacoonWW}. This generator
includes the complete
lowest-order matrix elements for $\eeffff$ for any four-fermion final
state. 
For the virtual corrections a DPA is used without any additional 
approximations. In particular, the exact four-fermion phase space is 
used throughout. The virtual corrections consist of factorizable
and non-factorizable contributions. The former are the ones that are
associated to either \PW-pair production or \PW-boson decay; the results
of \citeres{rcwprod1,rcwdecay1} are used in this part. The latter
comprise all corrections in which the subprocesses production and
decays do not proceed independently. Up to some simple supplements,
the virtual non-factorizable corrections can be read off from the
literature \cite{nfc1,nfc2}; we made use of the results of
\citere{nfc2}.  The real bremsstrahlung corrections are based on the
full matrix-element calculation for $\eeffffg$ described in
\citere{ee4fa}. More precisely, the minimal gauge-invariant subset
including all doubly-resonant contributions of the processes
$\eeWWffffg$, \ie the photon radiation from the CC11 subset, are
included.  By using the exact matrix elements for the real radiation,
we avoid problems in defining a DPA for semi-soft photons
($E_\gamma\sim\GW$) and, moreover, can include the leading logarithmic
corrections to non-doubly-resonant diagrams (background diagrams)
exactly.  The real corrections and the virtual corrections are matched
in such a way that all infrared singularities cancel exactly.  
The initial-state collinear singularities are regularized by retaining
a finite electron mass and factorized into lowest-order matrix element
and splitting functions. The collinear singularities connected to
final-state radiation are treated inclusively, \ie photons within
collinear cones around the final-state fermions are integrated over,
so that no logarithmic final-state fermion mass dependence survives.
All contributions have been implemented in two
programs, one of which uses the subtraction method
described in \citere{subtract}, the other one uses phase-space
slicing. All parts of the calculations have been performed in two
independent ways.
A detailed description of the calculation and the Monte Carlo
generator {\sc RacoonWW} will be published elsewhere \cite{De00}.
\bigskip

For the numerical results we used the fixed-width scheme and the
following parameters:
\beq
\begin{array}[b]{rlrl}
\GF =& 1.16637\times 10^{-5} \GeV^{-2}, \qquad&\alpha=&1/137.0359895, \\
\MW =& 80.35\GeV,& \GW =& 2.08699\ldots\GeV, \\
\MZ =& 91.1867\GeV,& \GZ =& 2.49471\GeV, \\
\Mt =& 174.17 \GeV,&\MH=& 150\GeV, \\
\Me =& 510.99907 \keV.
\end{array}
\eeq
The weak mixing angle is fixed by $\cw=\MW/\MZ$, $\sw^2=1-\cw^2$.
These parameters are over-complete but self-consistent. Instead
of $\alpha$ we use $\GF$ to parame\-trize the lowest-order matrix
element, \ie we use the effective coupling
\beq
\alpha_{\GF} = \frac{\sqrt{2}\GF\MW^2\sw^2}{\pi}
\eeq
in the lowest-order matrix element. This parameterization has the
advantage that all higher-order contributions associated with the
running of the electromagnetic coupling and the leading universal
two-loop $\Mt$-dependent 
corrections are correctly taken into account.
In the relative $\Oa$ corrections, on the other hand, we use $\al$,
since in the real corrections the scale of the real photon is zero.
The W-boson width given above is calculated including the electroweak
and QCD one-loop corrections with $\alpha_{\mathrm s}=0.119$.  We do
not include QCD corrections to the process $\eeWWffff$, and
initial-state radiation is only taken into account in $\Oa$.

In \reftas{table184}, \ref{table189}, and \ref{table200} we present
numbers for total cross sections without any cuts, for centre-of-mass
(CM) energies 184, 189, and $200\GeV$, respectively, based on 20
million events. In particular, we give the CC03 cross sections, \ie
the cross sections resulting from the signal diagrams only (defined in
the 't~Hooft--Feynman gauge). We also give numbers resulting from the
complete set of diagrams for those final states where this is possible
without cuts, \ie the CC11 class of processes.  In the considered
cases, the effects of the background diagrams are below $0.2\%$.  Note
that we treat the external fermions as massless. Therefore, the cross
sections for processes
not in CC11 class become singular if no cuts
are imposed for final-state electrons collinear to the beams and for
virtual photons splitting into $f\bar f$ pairs with small invariant
masses.
\begin{table}
$$
\begin{array}{l@{\qquad}c@{\qquad}c@{\qquad}c@{\qquad}c}
\hline
\mbox{final state} &
\mbox{CC03 Born} &  \mbox{full Born} & 
\mbox{CC03 corrected} & \mbox{full corrected} \nl
\hline
\nu_\mu \mu^+ \tau^-\bar\nu_\tau &
210.26(5)  &  210.57(5) & 182.21(11) &  182.53(11)  
\nl
\nu_\mu\mu^+\Pd\Pubar   & 
 630.8(2)  &  631.8(2)  & 546.5(3)   &   547.4(3)   
\nl
\Pu\Pdbar\Ps\Pcbar & 
 1892.3(5) &  1895.4(5) & 1637.7(7)  &  1640.7(7)   
\nl
\mbox{total} &
17031(3) & & 14749(5) \nl 
\hline
\end{array}
$$
\caption{Total cross sections in fb for $\eeWWffff$ without cuts 
for various final states at $184\GeV$}
\label{table184}
\vspace*{1em}
$$
\begin{array}{l@{\qquad}c@{\qquad}c@{\qquad}c@{\qquad}c}
\hline
\mbox{final state} &
\mbox{CC03 Born} &  \mbox{full Born} & 
\mbox{CC03 corrected} & \mbox{full corrected} \nl
\hline
\nu_\mu \mu^+ \tau^-\bar\nu_\tau &
216.07(6)  &  216.39(6) & 190.64(11)&  190.96(11)  
\nl
\nu_\mu\mu^+\Pd\Pubar   & 
 648.2(2)  &  649.2(2)  & 571.6(3)  &   572.6(3)   
\nl
\Pu\Pdbar\Ps\Pcbar & 
 1944.7(5) &  1947.7(5) & 1713.4(8) &  1716.4(8)   
\nl
\mbox{total} &
17501(3) & & 15429(5) \nl  
\hline
\end{array}
$$
\caption{Total cross sections in fb for $\eeWWffff$ without cuts 
for various final states at $189\GeV$}
\label{table189}
\vspace*{1em}
$$
\begin{array}{l@{\qquad}c@{\qquad}c@{\qquad}c@{\qquad}c}
\hline
\mbox{final state} &
\mbox{CC03 Born} &  \mbox{full Born} & 
\mbox{CC03 corrected} & \mbox{full corrected} \nl
\hline
\nu_\mu \mu^+ \tau^-\bar\nu_\tau &
219.82(6)  &  220.06(6) & 199.74(12) &  199.98(12)  
\nl
\nu_\mu\mu^+\Pd\Pubar   & 
 659.5(2)  &  660.2(2)  & 598.9(3)   &   599.7(3)   
\nl
\Pu\Pdbar\Ps\Pcbar & 
 1978.4(5) &  1980.8(5) & 1795.0(9)  &  1797.4(9)   
\nl
\mbox{total} &
17806(3) & & 16164(5) \nl    
\hline
\end{array}
$$
\caption{Total cross sections in fb for $\eeWWffff$ without cuts 
for various final states at $200\GeV$}
\label{table200}
\end{table}
The shown corrections are typically $-13\%$, $-12\%$, and $-9\%$ at
184, 189, and $200\GeV$, respectively. The numbers in parentheses are
estimates of the Monte Carlo integration errors.  The numbers for the
total \PW-pair production cross section in the last row of the tables
directly result from the other rows by multiplying these with the
number of equivalent channels and adding them up.

Next, we study various angular distributions. Here we restrict
ourselves to the $\nu_\mu\mu^+\Pd\Pubar$ final state and to
$\sqrt{s}=200\GeV$.  Because of our treatment of mass singularities
(see \citere{De00} for details) it is necessary to combine photons
that are collinear to the incoming or outgoing fermions appropriately
with these fermions in order to obtain well-defined finite
distributions.
To this end we introduce the following recombination and cut procedure
which proceeds in three steps:
\newcommand{\recomb}{{\mathrm{rec}}}%
\begin{enumerate}
\item All photons within a cone of 5 degrees around the beams are
  treated as invisible, \ie their momenta are disregarded when
  calculating angles, energies,
  and invariant masses.
\item Next, the invariant masses of the photon with 
each of the charged final-state fermions are calculated. If the
smallest one is smaller than $M_\recomb$, the photon is combined with
the corresponding fermion, \ie the momenta of the photon and the
fermion are added and associated with the momentum of the fermion,
and the photon is discarded.
\item Finally, all events are discarded 
  in which one of the
  final-state fermions is within a cone of 10 degrees around the
  beams.  No other cuts are applied.
\end{enumerate}
We consider the cases of a tight recombination cut $M_\recomb= 5\GeV$
and of a loose recombination cut $M_\recomb= 25\GeV$.
In the following observables, the momenta of the \PW~bosons are always
defined by the sum of the momenta of the two corresponding decay
fermions after the eventual recombination with the photon.

The results for the distributions have been obtained from 50 million
events. In the following figures we always show on the left-hand side
the absolute distributions in lowest order and including the
corrections for the recombination cut $M_\recomb=5\GeV$,
and on the
right-hand side the corresponding relative corrections for the two
recombination cuts $M_\recomb=5\GeV$ and $M_\recomb=25\GeV$.

\begin{figure}
\centerline{%
\setlength{\unitlength}{1cm}
\begin{picture}(7.9,8.2)
\put(0,0){\includegraphics{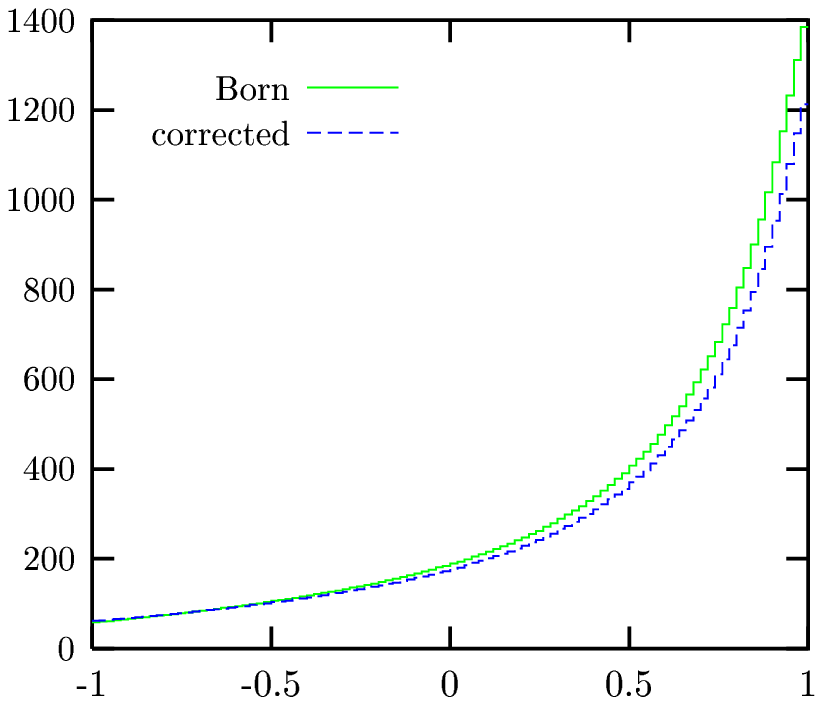}}
\put(0.1,7.3){\makebox(1,1)[l]{$\frac{\rd \si}{\rd \cos\theta_{\PW}}\ 
\left[\fb\right]$}}
\put(4.0,-0.2){\makebox(1,1)[c]{$\cos\theta_{\PW}$}}
\end{picture}%
\begin{picture}(7.9,8.2)
\put(0.5,7.3){\makebox(1,1)[c]{$\de\ [\%]$}} 
\put(4.0,-0.2){\makebox(1,1)[c]{$\cos\theta_{\PW}$}}
\put(0,0){\includegraphics{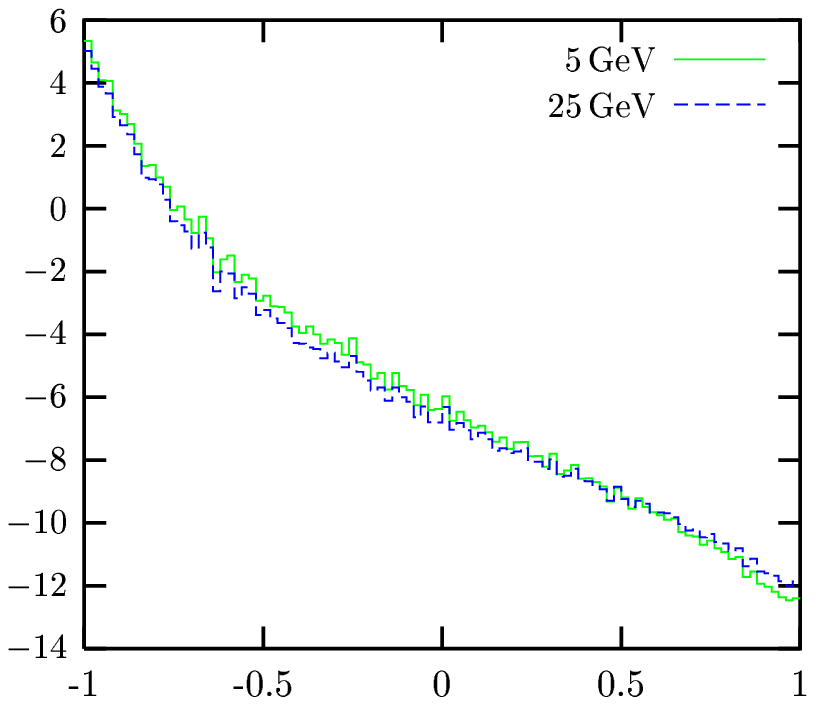}}
\end{picture}
}
\caption{Production-angle distribution for
  $\Pep\Pem\to\nu_\mu\mu^+\Pd\Pubar$ and $\protect\sqrt{s}=200\GeV$}
\label{fi:prod_angle}
\end{figure}

In \reffi{fi:prod_angle} we show the distribution of events in the
angle between the $\PWp$ and the incoming $\Pep$. Apart from the
normalization effects, a distortion of the distribution occurs. 
This is mainly due to hard-photon emission from the initial state,
which boosts the CM-system of the W bosons and causes a migration of
events from regions with large cross section in the CM system to
regions with small cross section in the laboratory system. This effect
is also visible in the following distributions.  The production-angle
distribution hardly depends on the recombination scheme, as expected.
An increase of the recombination cut leads to a small redistribution
of events from the backward to the forward direction.

\begin{figure}
\centerline{%
\setlength{\unitlength}{1cm}
\begin{picture}(7.9,8.2)
\put(0,0){\includegraphics{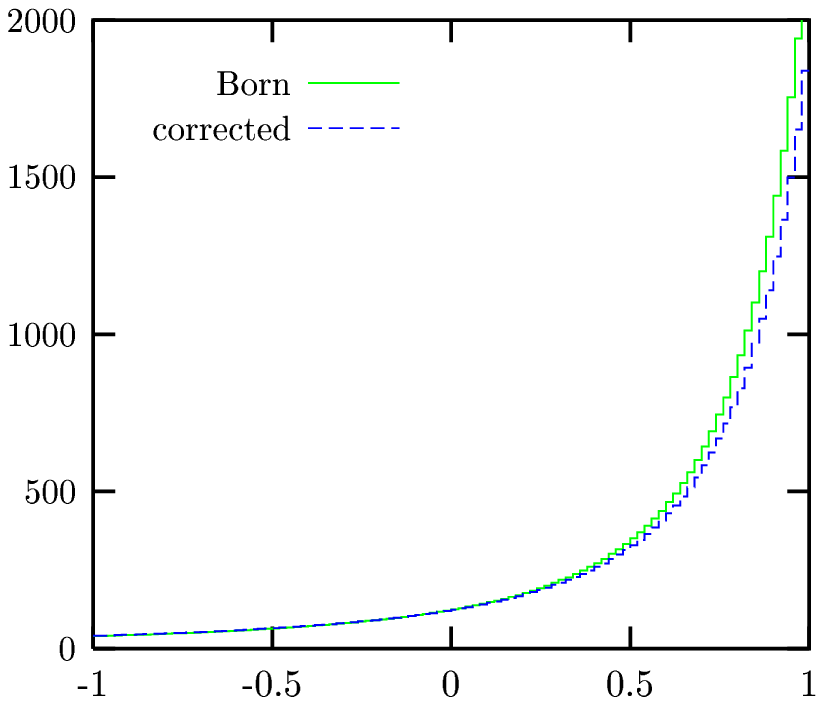}}
\put(0.1,7.3){\makebox(1,1)[l]{$\frac{\rd \si}{\rd \cos\theta_{\PWp\mu^+}}\ 
\left[\fb\right]$}}
\put(4.0,-0.2){\makebox(1,1)[c]{$\cos\theta_{\PWp\mu^+}$}}
\end{picture}%
\begin{picture}(7.9,8.2)
\put(0.5,7.3){\makebox(1,1)[c]{$\de\ [\%]$}} 
\put(4.0,-0.2){\makebox(1,1)[c]{$\cos\theta_{\PWp\mu^+}$}}
\put(0,0){\includegraphics{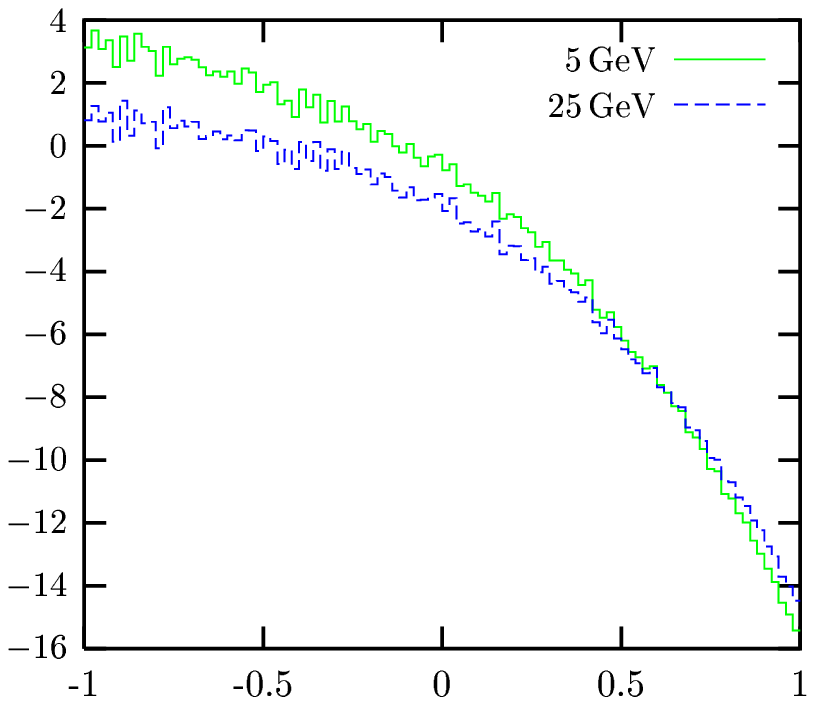}}
\end{picture}
}
\caption{Decay-angle distribution for
  $\Pep\Pem\to\nu_\mu\mu^+\Pd\Pubar$ and $\protect\sqrt{s}=200\GeV$}
\label{fi:decay_angle}
\end{figure}

The distribution of events in the angle between the $\PWp$ and the
outgoing $\mu^+$ is presented in \reffi{fi:decay_angle}.
In this case we find a sizeable dependence on the recombination mass
$M_\recomb$ for large decay angles, where the cross section is small.
This originates from the fact 
that the recombination of a fermion with a photon 
parallel to this fermion
decreases the angle between the fermion
and the W~boson from which the fermion results.

\begin{figure}
\centerline{%
\setlength{\unitlength}{1cm}
\begin{picture}(7.9,8.2)
\put(0,0){\includegraphics{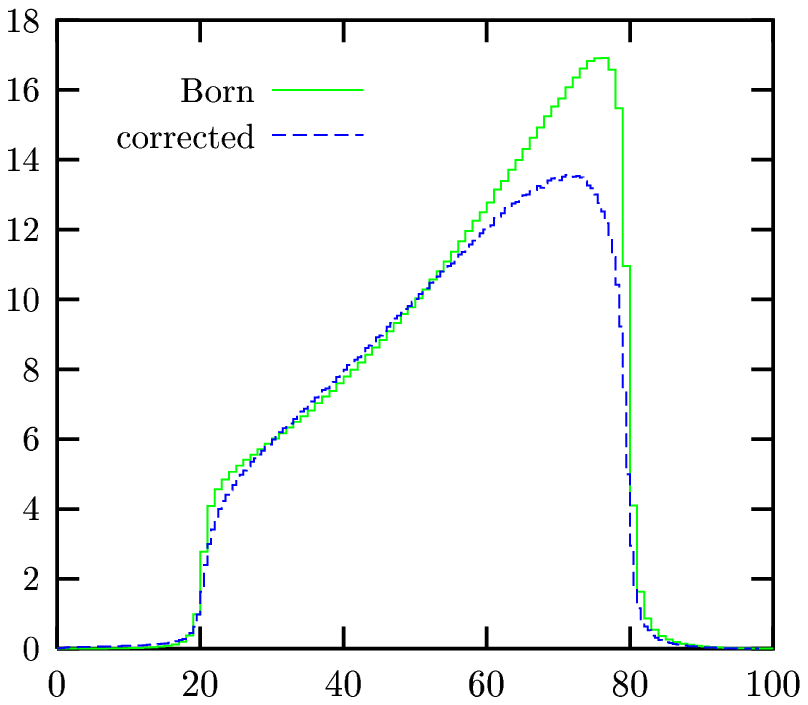}}
\put(0.1,7.3){\makebox(1,1)[l]{$\frac{\rd \si}{\rd E_\mu}\ 
\left[\frac{\fb}\GeV\right]$}}
\put(4.0,-0.2){\makebox(1,1)[c]{$E_\mu\ [\mathrm{GeV}]$}}
\end{picture}%
\begin{picture}(7.9,8.2)
\put(0.5,7.3){\makebox(1,1)[c]{$\de\ [\%]$}} 
\put(4.0,-0.2){\makebox(1,1)[c]{$E_\mu \ [\mathrm{GeV}]$}}
\put(0,0){\includegraphics{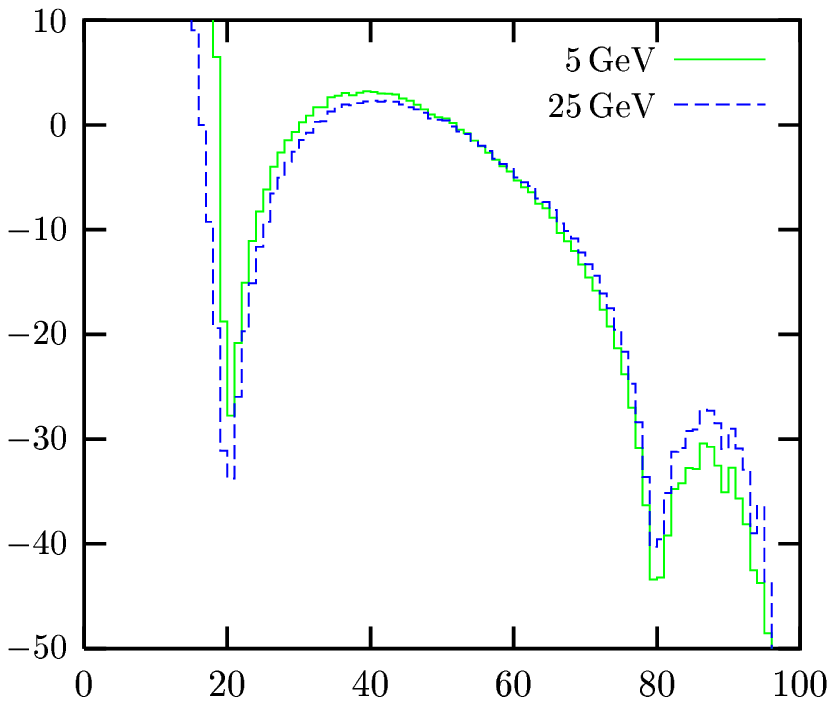}}
\end{picture}
}
\caption{Muon-energy distribution for
  $\Pep\Pem\to\nu_\mu\mu^+\Pd\Pubar$ and $\protect\sqrt{s}=200\GeV$}
\label{fi:mu_energy}
\end{figure}

The distribution of events in the energy $E_\mu$ of the 
outgoing $\mu^+$ is depicted in \reffi{fi:mu_energy}.
In the on-shell approximation it would be restricted between
$20.2\GeV<E_\mu<79.8\GeV$. 
Outside this region, the corrections
calculated in DPA are not reliable. 
The recombination of a photon with the
muon increases the muon energy. Consequently, an increase of the 
recombination mass $M_\recomb$ shifts the distribution to larger muon
energies as can be seen in the relative corrections.

\begin{figure}
\centerline{%
\setlength{\unitlength}{1cm}
\begin{picture}(7.9,8.2)
\put(0,0){\includegraphics{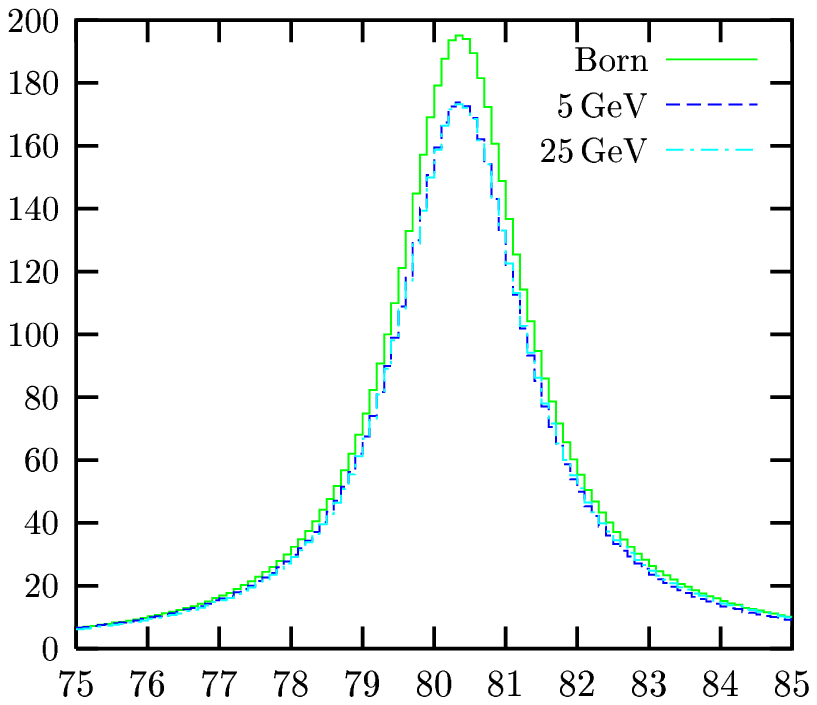}}
\put(0.1,7.3){\makebox(1,1)[l]{$\frac{\rd \si}{\rd M_{\mu\nu_\mu}}\ 
\left[\frac{\fb}\GeV\right]$}}
\put(4.0,-0.2){\makebox(1,1)[c]{$M_{\mu\nu_\mu}\ [\mathrm{GeV}]$}}
\end{picture}%
\begin{picture}(7.9,8.2)
\put(0.5,7.3){\makebox(1,1)[c]{$\de\ [\%]$}} 
\put(4.0,-0.2){\makebox(1,1)[c]{$M_{\mu\nu_\mu}\ [\mathrm{GeV}]$}}
\put(0,0){\includegraphics{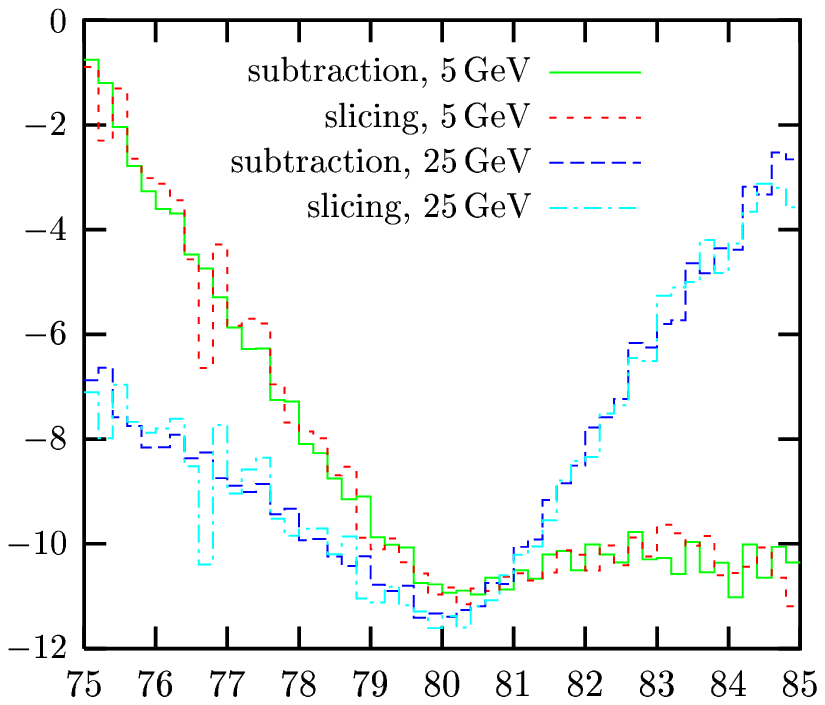}}
\end{picture}
}
\caption{Invariant-mass distribution of the lepton pair for
  $\Pep\Pem\to\nu_\mu\mu^+\Pd\Pubar$ and $\protect\sqrt{s}=200\GeV$}
\label{fi:munu_invmass}
\end{figure}

\begin{figure}
\centerline{%
\setlength{\unitlength}{1cm}
\begin{picture}(7.9,8.2)
\put(0,0){\includegraphics{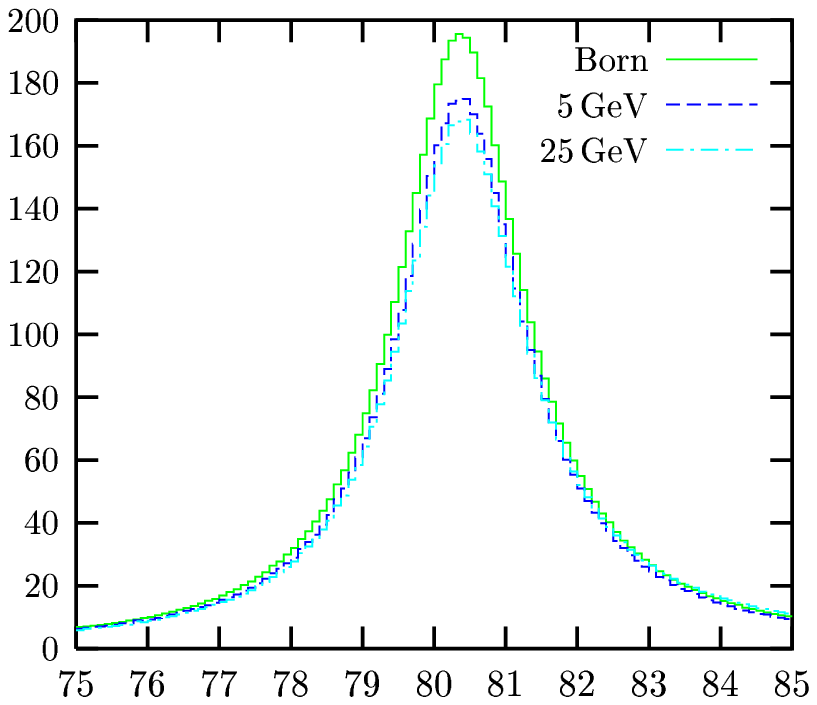}}
\put(0.1,7.3){\makebox(1,1)[l]{$\frac{\rd \si}{\rd M_{\Pd\Pu}}\ 
\left[\frac{\fb}\GeV\right]$}}
\put(4.0,-0.2){\makebox(1,1)[c]{$M_{\Pd\Pu}\ [\mathrm{GeV}]$}}
\end{picture}%
\begin{picture}(7.9,8.2)
\put(0.5,7.3){\makebox(1,1)[c]{$\de\ [\%]$}} 
\put(4.0,-0.2){\makebox(1,1)[c]{$M_{\Pd\Pu}\ [\mathrm{GeV}]$}}
\put(0,0){\includegraphics{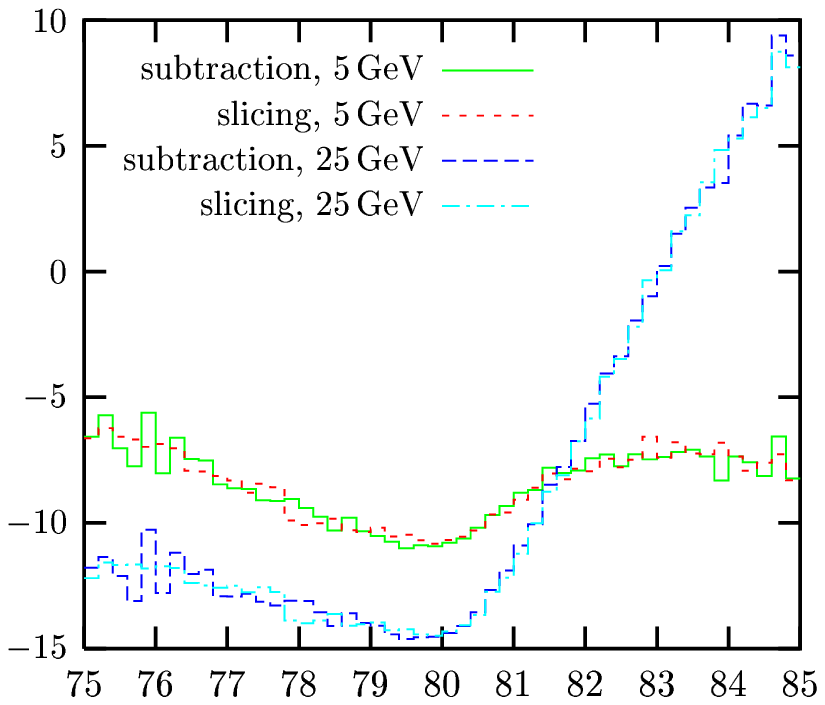}}
\end{picture}
}
\caption{Invariant-mass distribution of the quark pair for
  $\Pep\Pem\to\nu_\mu\mu^+\Pd\Pubar$ and $\protect\sqrt{s}=200\GeV$}
\label{fi:ud_invmass}
\end{figure}

Finally, in \reffis{fi:munu_invmass} and \ref{fi:ud_invmass} we show 
the distributions of events in the invariant masses of the final-state
lepton pair, $M_{\mu\nu_\mu}$, and of the final-state quark pair,
$M_{\Pd\Pu}$. The dependence on the recombination cut is
sizeable everywhere.
The results for the invariant-mass distributions can be understood as
follows. For small recombination cuts, in most of the events the
W~bosons are defined from the decay fermions only. If a photon is
emitted from the decay fermions
and not recombined, the invariant mass of the fermions is smaller than
the one of the decaying W~boson. This leads to an enhancement of the
distribution for invariant masses below the \PW~resonance. This effect
becomes smaller with increasing recombination mass.  The enhancement
is proportional to the squared charges of the final-state fermion, \ie
it is largest for the leptonic invariant mass. On the other hand, if
the recombination mass gets large,
the probability increases that the recombined fermion momenta receive
contributions from photons that are radiated during the W-production
subprocess or from the decay fermions of the other W~boson.
This leads to positive corrections above the considered \PW~resonance.
The effect is larger for the hadronic invariant mass since in this
case, two decay fermions (the two quarks) can be combined with the
photon. The effect of the squared charges of the final-state fermions
is marginal in this case because the contribution of initial-state
fermions dominates.

The distortion of the W~invariant-mass distributions is of particular
interest for the reconstruction of the W-boson mass $\MW$ from the decay
products. In order to illustrate the impact of the corrections on the
determination of $\MW$, we fit the simple Breit--Wigner distribution 
\beq
\biggl(\frac{\rd\sigma}{\rd M^2}\biggr) =
\frac{\mbox{const.}}{(M^2-\MWfit^2)^2+\MWfit^2 \GWfit^2}
\eeq
to the W~line shapes shown in \reffis{fi:munu_invmass} and
\ref{fi:ud_invmass}, treating $\MWfit$ and $\GWfit$ (as well as the
normalization constant) as free fit parameters. 
We determine the fitted \PW-boson masses from the predictions
resulting from the lowest-order CC03 diagrams, $\MWfit^{\born,\mathrm{CC03}}$,
from the complete lowest-order diagrams, $\MWfit^{\born}$,  and from
the fully corrected predictions, $\MWfit^{\corr}$. In addition we give
the mass shift resulting from the corrections
$\Delta\MWfit^{\corr} = \MWfit^{\corr} - \MWfit^{\born}$.
The results of these
fits, which are contained in \refta{tablemw}, show that the fitted 
W-boson mass changes at the order of some $10\MeV$ if the corrections are
included. From the discussion of the line-shape distortion above it is
clear that this mass shift is more positive if more photons are
recombined. The results also illustrate that the fit results vary
at the order of some $10\MeV$ for different fit ranges in $M$.
\begin{table}
$$
\begin{array}{c@{\qquad}c@{\qquad}c@{\qquad}c@{\qquad}c@{\qquad}c}
\hline
& M_\recomb &
\MWfit^{\born,\mathrm{CC03}} & \MWfit^{\born} &  
\MWfit^{\corr} & \Delta\MWfit^{\corr} \nl
& [\mathrm{GeV}] &
[\mathrm{GeV}] & [\mathrm{GeV}] &  
[\mathrm{GeV}]  & [\mathrm{MeV}] \nl
\hline
\PWp\to\nu_\mu \mu^+ &  5 & 80.366 & 80.365 & 80.363 &  -2 \nl
                     &    & 80.372 & 80.371 & 80.365 &  -5 \nl
                     & 25 & 80.366 & 80.365 & 80.374 &  +9 \nl
                     &    & 80.372 & 80.371 & 80.385 & +14 \nl
\hline
\PWm\to\Pd\Pubar     &  5 & 80.365 & 80.364 & 80.378 & +14 \nl
                     &    & 80.371 & 80.370 & 80.384 & +14 \nl
                     & 25 & 80.365 & 80.364 & 80.397 & +32 \nl
                     &    & 80.371 & 80.370 & 80.415 & +46 \nl
\hline
\end{array}
$$
\caption{Results for $\MWfit$ for a Breit--Wigner fit to the 
invariant-mass distributions shown in \reffis{fi:munu_invmass} and 
\ref{fi:ud_invmass}, using the fit ranges 
$78.3\GeV < M < 82.3\GeV$ (upper values) and
$76.3\GeV < M < 84.3\GeV$ (lower values) }
\label{tablemw}
\end{table}

We have also considered the case where each photon is recombined with
the nearest fermion, \ie $M_\recomb>\sqrt{s}$. In the distributions of
the production angle, the decay angle, and the muon energy this
increase of the recombination cut leads to the expected small changes
of the corrections. For the invariant-mass distributions, the case of
complete recombination differs from the case $M_\recomb=25\GeV$ only
by a slight decrease of the relative corrections. This is due to the
fact that the recombination of a photon that forms an invariant mass
of at least $25\GeV$ with charged fermions
shifts the events by at least
$3.8 \GeV$ for the leptonically-decaying \PW~boson and $7.4\GeV$ for the
hadronically-decaying \PW~boson. Such a shift leads to an increase of
the corrections only outside the range shown in
\reffis{fi:munu_invmass} and \ref{fi:ud_invmass}. Since there is no
change in the shape of the distribution near the resonance, the
results for the invariant-mass fit are essentially equivalent.

In the plots on the right-hand 
side in \reffis{fi:munu_invmass} and
\ref{fi:ud_invmass} we have included the results for phase-space
slicing and for the subtraction method. 
The agreement between these
results demonstrates the correctness of our implementations and gives 
another estimate on the size of the integration errors.
\bigskip

In summary, we have constructed the event generator {\sc RacoonWW} for
$\eeWWffff(+\gamma)$ which includes the complete $\Oa$ electroweak
corrections in double-pole approximation. With this generator we have
calculated the total cross sections, including the CC03 cross
sections, and various distributions of experimental relevance for
typical LEP2 energies.  The detailed numerical discussion of the
corrections, in particular, illustrates the importance of the issue of
photon recombination.  More details of the {\sc RacoonWW} approach and
further results, including a comparison to results of other authors,
will be presented elsewhere \cite{De00}.


\begin{thebibliography}{99}
\frenchspacing
\newcommand{\ap}[3]{{\sl Ann.~Phys.} {\bf #1} (19#2) #3}
\newcommand{\app}[3]{{\sl Acta.~Phys.~Pol.} {\bf #1} (19#2) #3}
\newcommand{\zp}[3]{{\sl Z.~Phys.} {\bf #1} (19#2) #3}
\newcommand{\np}[3]{{\sl Nucl.~Phys.} {\bf #1} (19#2) #3}
\newcommand{\pl}[3]{{\sl Phys.~Lett.} {\bf #1} (19#2) #3}
\newcommand{\pr}[3]{{\sl Phys.~Rev.} {\bf #1} (19#2) #3}
\newcommand{\prl}[3]{{\sl Phys.~Rev.~Lett.} {\bf #1} (19#2) #3}
\newcommand{\fp}[3]{{\sl Fortschr.~Phys.} {\bf #1} (19#2) #3}
\newcommand{\jp}[3]{{\sl J.~Phys.} {\bf #1} (19#2) #3}
\newcommand{\cpc}[3]{{\sl Comput.~Phys.~Commun.} {\bf #1} (19#2) #3}
\newcommand{\ijmp}[3]{{\sl Int.~J.~Mod.~Phys.} {\bf #1} (19#2) #3}
\newcommand{\nim}[3]{{\sl Nucl.~Instr.~Meth.} {\bf #1} (19#2) #3}
\newcommand{\nc}[3]{{\sl Nuovo Cimento} {\bf #1} (19#2) #3}
\newcommand{\vj}[4]{{\sl #1} {\bf #2} (19#3) #4}

\bibitem{rcwprod1}
M.~B\"ohm et al., \np{B304}{88}{463}.

\bibitem{rcwprod2}
J. Fleischer, F. Jegerlehner and M. Zra\l ek, \zp{C42}{89}{409}.

\bibitem{rcwdecay1}
A.~Denner and T. Sack, \zp{C46}{90}{653}.

\bibitem{rcwdecay2}
D.Yu. Bardin, S. Riemann and T. Riemann,  \zp{C32}{86}{121};\\
F. Jegerlehner, \zp{C32}{86}{425}.

\bibitem{Ae94}
A. Aeppli, G.J. van Oldenborgh and D. Wyler, \np{B428}{94}{126}.

\bibitem{ja97}
S.~Jadach et al., \pl{B417}{98}{326}.

\bibitem{ja99}
S.~Jadach et al., hep-ph/9907436.

\bibitem{Be98}
W. Beenakker,  F.A. Berends and A.P. Chapovsky, \np{B548}{99}{3}.

\bibitem{nfc1}
K. Melnikov and O.I. Yakovlev, \np{B471}{96}{90}; \\
W. Beenakker,  A.P. Chapovsky and F.A. Berends,
\pl{B411}{97}{203} and \np{B508}{97}{17}.

\bibitem{nfc2}
A. Denner, S. Dittmaier and M. Roth, \np{B519}{98}{39} and
\pl{B429}{98}{145}.

\bibitem{ee4fa}
A. Denner, S. Dittmaier, M. Roth and D. Wackeroth, 
\np{B560}{99}{33}.

\bibitem{subtract} 
S.~Dittmaier, BI-TP 99/09, hep-ph/9904440, 
to appear in {\sl Nucl.~Phys.} {\bf B};\\
M. Roth, dissertation ETH Z\"urich No.~13363, 1999.

\bibitem{De00}
A. Denner, S. Dittmaier, M. Roth and D. Wackeroth, in preparation.












\end{thebibliography}
\end{document}